\documentclass[twocolumn]{emulateapj}

\usepackage{amsmath,color}

\slugcomment{\textbf{Astrophysical Journal Letters, in press}}

\shorttitle{Interstellar Comet 2I/Borisov}
\shortauthors{Kim et al.}

\begin{document}

\title{Coma Anisotropy and the Rotation Pole of \\Interstellar Comet 2I/Borisov}

\author{Yoonyoung Kim$^1$,
David Jewitt$^{2,3}$,
Max Mutchler$^4$,
Jessica Agarwal$^1$,
Man-To Hui$^5$ and
Harold Weaver$^6$  
}

\affil{$^1$ Max Planck Institute for Solar System Research, 37077 G\"ottingen, Germany\\
$^2$ Department of Earth, Planetary and Space Sciences,
UCLA, Los Angeles, CA 90095-1567\\
$^3$ Department of Physics and Astronomy,
UCLA, Los Angeles, CA 90095-1547\\
$^4$ Space Telescope Science Institute, Baltimore, MD 21218 \\
$^5$ Institute for Astronomy, University of Hawaii,  Honolulu, Hawaii 96822\\
$^6$ The Johns Hopkins University Applied Physics Laboratory,  Laurel, Maryland 20723
}

\email{kimy@mps.mpg.de}

\begin{abstract} 

Hubble Space Telescope observations of interstellar comet 2I/Borisov near perihelion show the ejection of large ($\gtrsim$100 $\mu$m) particles at $\lesssim$9 m s$^{-1}$ speeds, with estimated mass-loss rates of $\sim$35~kg~s$^{-1}$. 
The total mass loss from comet Borisov corresponds to loss of a surface shell on the nucleus only $\sim$0.4~m thick.  This shell is thin enough to be susceptible to past chemical processing in the interstellar medium by cosmic rays, meaning that the ejected materials cannot necessarily be considered as pristine.  
Our high-resolution images reveal persistent asymmetry in the dust coma, best explained by a thermal lag on the rotating nucleus causing peak mass loss to occur in the comet nucleus afternoon.  
In this interpretation, the nucleus rotates with an obliquity of 30$\degr$ (pole direction RA = 205$\degr$ and Dec. = 52$\degr$).
The subsolar latitude varied from $-35\degr$ (southern solstice) at the time of discovery to $0\degr$ (equinox) in 2020 January, suggesting the importance of seasonal effects.
Subsequent activity likely results from  regions freshly activated as the northern hemisphere is  illuminated for the first time.\\

\end{abstract}


\section{INTRODUCTION}

Comet 2I/Borisov (formerly C/2019 Q4 and, hereafter, ``2I'') is the first known interstellar comet (Borisov 2019) and only the second interstellar object  identified in the solar system.  It likely originated in the protoplanetary disk of another star.  Scientific interest in 2I centers on characterizing this object, both to compare it with the first interstellar body, 1I/'Oumuamua, and with the long- and short-period comet populations of the solar system.  Early ground-based observations established that 2I was continuously active, with a coma consisting of $\sim$100 $\mu$m sized, slowly-moving grains and optical colors consistent with those of ``normal'' solar system comets (Jewitt \& Luu 2019, Guzik et al.~2020, Ye et al.~2020).  Observations using the Hubble Space Telescope (HST, Jewitt et al.~2020) revealed a small nucleus, having a radius, $r_n$, in the range 0.2 $\le r_n \le$ 0.5 km (a smaller nucleus would show non-gravitational acceleration larger than measured while a larger nucleus would be discernible in the high resolution HST surface brightness profile).  Spectroscopic observations established the presence of CN gas (Fitzsimmons et al.~2019), with water production rates, inferred from detection of cometary oxygen, of  $dM/dt \sim$ (20~$\pm$~5) kg s$^{-1}$ at $r_H$ = 2.38 AU  (McKay et al.~2020).
The upper limits reported for C$_2$ suggest that 2I is carbon-chain depleted (Opitom et al.~2019, Kareta et al.~2020).
Carbon monoxide, CO, is abundant, with a post-perihelion production rate near 20 to 40 kg s$^{-1}$ (Bodewits et al.~2020, Cordiner et al.~2020).

Here, we report continued observations of 2I taken with the HST,  focusing on a  detailed morphological examination of the coma of this intriguing object.  
We combine the highest-resolution optical data with a sophisticated Monte Carlo model of cometary dust dynamics to quantify the properties of the dust coma and its  underlying nucleus source.

\section{OBSERVATIONS}

Observations with the HST were taken under General Observer programs 16009 and 16041. 
We used the UVIS channel of the WFC3 camera with the broadband F350LP filter (effective wavelength $\sim$5846\AA, FWHM $\sim$4758\AA) in order to maximize the signal-to-noise ratios in the data. The image scale and the field of view are 0.04\arcsec~pixel$^{-1}$ and 80\arcsec$\times$80\arcsec, respectively, where the comet is centrally located.
The earliest observations (UT 2019 October 12) were obtained from four closely-spaced HST orbits (Jewitt et al.~2020), while observations on the remaining dates were each obtained from a single HST orbit within each of which  we obtained six exposures of 210--260 s duration.
A journal of observations is given in Table \ref{geometry}.

Our data, taken between UT 2019 October 12 and UT 2020 January 29, provide a 3.5 month window on the activity of 2I around perihelion (which occurred on UT 2019 December 09 at $q$ = 2.007 AU) and offer a range of viewing perspectives useful for characterizing the dust.  Special observations  were targeted on UT 2020 January 29 as the Earth passed through the projected orbital plane of 2I (the out-of-plane angle was 0.2\degr).  This viewing geometry provides a particularly powerful constraint on the out-of-plane distribution of dust and, hence, on its ejection velocity.

\begin{deluxetable*}{lcccrccccr}
\tablecaption{Observing Geometry 
\label{geometry}}
\tablewidth{0pt}
\tablehead{ \colhead{UT Date and Time}   & \colhead{DOY\tablenotemark{a}} & $\Delta T_p$\tablenotemark{b} & \colhead{$r_H$\tablenotemark{c}}  & \colhead{$\Delta$\tablenotemark{d}} & \colhead{$\alpha$\tablenotemark{e}}   & \colhead{$\theta_{-\odot}$\tablenotemark{f}} &   \colhead{$\theta_{-V}$\tablenotemark{g}}  & \colhead{$\delta_{\oplus}$\tablenotemark{h}}   }
\startdata

2019 Oct 12  13:44 - 20:42   & 285 &  -57     &  2.370 &  2.784  & 20.4 & 292.6   & 330.1 & -13.6 \\
2019 Nov 16  04:46 - 05:24   & 320 &  -22     &  2.067 &  2.205  & 26.5 & 289.2   & 330.9 & -17.6 \\
2019 Dec 09  12:03 - 12:41   & 343 &  1     &  2.006 &  1.988  & 28.5 & 289.9   & 327.8 & -15.6 \\
2020 Jan 03  03:18 - 03:56   & 368 &  26     &  2.085 &  1.941  & 28.0 & 294.5   & 319.3 & -9.0 \\
2020 Jan 29  11:34 - 12:07   & 394 &  52     &  2.313 &  2.058  & 25.2 & 305.0   & 304.4 & 0.2 
\enddata

\tablenotetext{a}{Day of Year, UT 2019 January 01 = 1}
\tablenotetext{b}{Number of days from perihelion (UT 2019-Dec-08 = DOY 342).}
\tablenotetext{c}{Heliocentric distance, in AU}
\tablenotetext{d}{Geocentric distance, in AU}
\tablenotetext{e}{Phase angle, in degrees}
\tablenotetext{f}{Position angle of the projected anti-Solar direction, in degrees}
\tablenotetext{g}{Position angle of the projected negative heliocentric velocity vector, in degrees}
\tablenotetext{h}{Angle of Earth above the orbital plane, in degrees\\}

\end{deluxetable*}

\section{RESULTS}

\subsection{Morphology}
\begin{figure*}
\epsscale{0.9}
\plotone{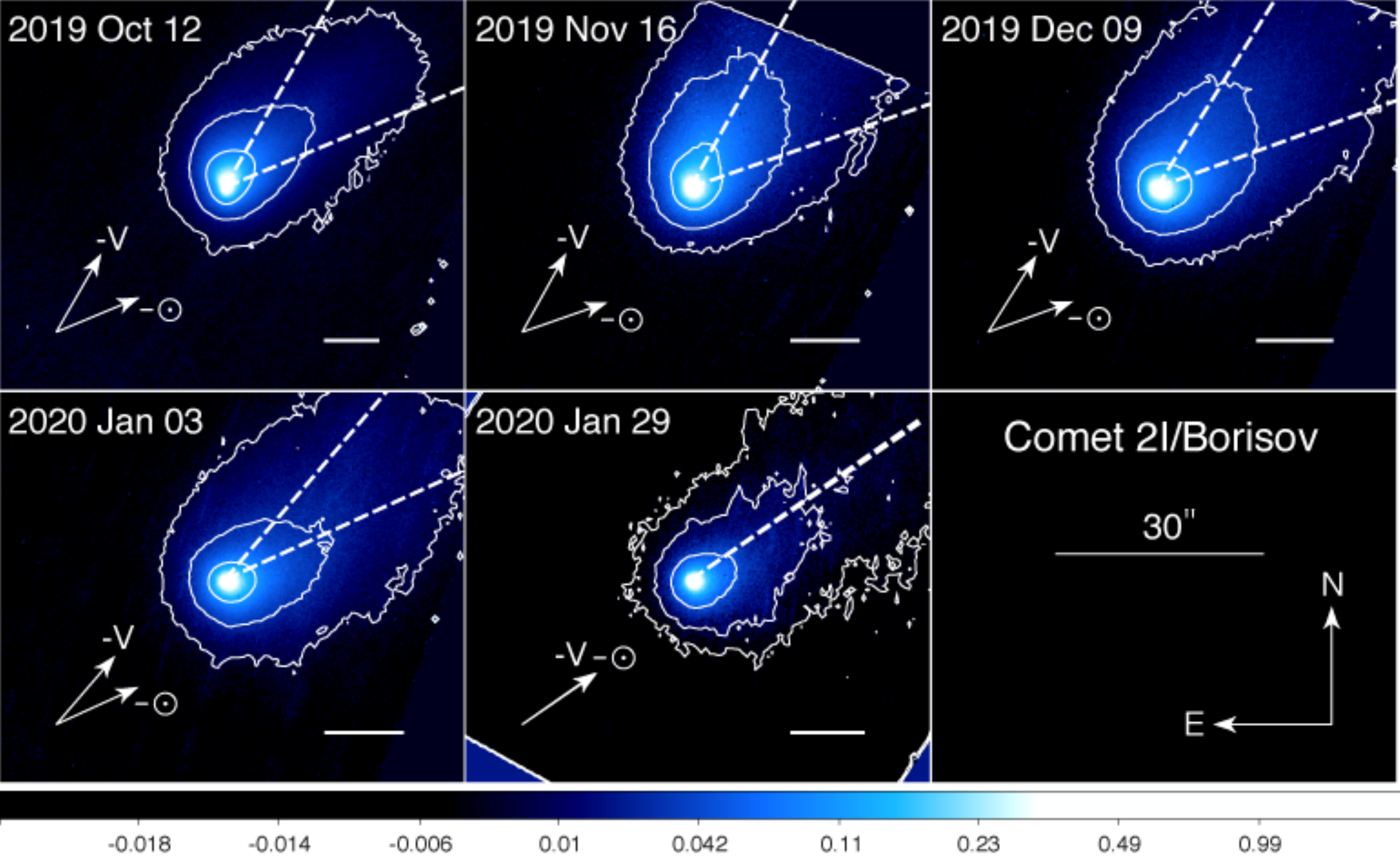}
\caption{Composite HST images of comet 2I/Borisov marked with UT dates of observation.
A color bar, the cardinal directions and the projected anti-solar direction ($-\odot$) and the negative heliocentric velocity vector ($-V$) are indicated.
Isophotal contours and extended $-\odot$ and $-V$ vectors (dashed lines) are overlaid to highlight a slight asymmetry in the inner coma.  A 16,000~km linear scale bar is shown for each date of observation.\\ \label{images}}
\end{figure*}

Composite images of 2I for each date of observation are shown in Figure \ref{images}.
Observations on all dates show dust extending in  directions between the projected anti-solar direction ($-\odot$) and the negative heliocentric velocity vector ($-V$), and show a slight asymmetry in the inner coma.  The October 12 composite image shows the visible edge of the tail at about 40\arcsec~from the nucleus, corresponding to a sky-plane distance $\ell \sim 8\times10^4$ km.
The tail observed between November 16 and January 29 extends beyond the WFC3 field of view ($>$40\arcsec).

The motion of cometary dust particles is controlled by $\beta$, the ratio of radiation pressure acceleration to solar gravity.  $\beta$ is a function of particle size, approximately given by $\beta \sim a^{-1}$, where $a$ is the particle radius expressed in microns.  For each epoch of observation, we first computed syndyne trajectories, defined as the loci of particles of a given $\beta$ ejected at different times with zero ejection velocity (Finson \& Probstein 1968).
The coma direction on October 12 is best-matched by the syndyne having $\beta \sim$  0.01 (or size $a \sim$ 100 $\mu$m), consistent with earlier reports (Jewitt \& Luu 2019; Jewitt et al.~2020).
However, we found that the direction of the dust coma observed in November and December does not coincide with the syndyne having $\beta \sim$  0.01, indicating that the coma particle size cannot be characterized by a single value.
The failure of syndyne trajectories to match the coma direction suggests that particles are ejected anisotropically.
In addition, an asymmetry in the January 29 composite image (in-plane observations) is evidence that the out-of-plane ejection was preferentially to the south.

\begin{figure}
\epsscale{0.93}
\plotone{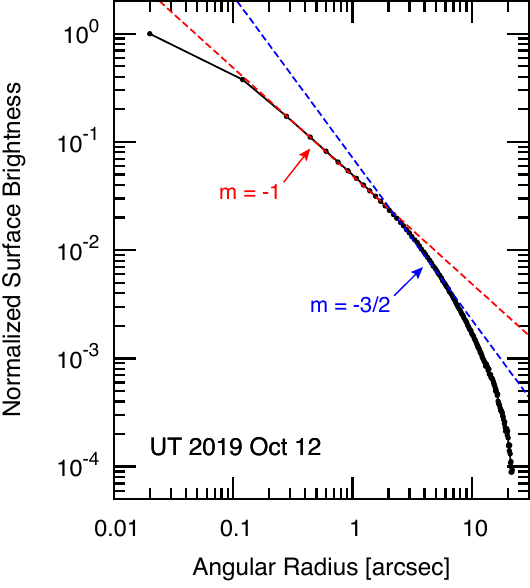}
\caption{Surface brightness profile of 2I observed on UT 2019 October 12. 
Lines indicate logarithmic surface brightness gradients $m$ = -1 (red) and $m$ = -3/2 (blue).
Background signal was determined in the annulus with inner and outer radii of 22\arcsec~and 26\arcsec, respectively.
At large angles $\gtrsim$ 7\arcsec, uncertainties in the background subtraction become significant.\\ \label{sbr}}
\end{figure}

\smallskip
\subsection{Radial Surface Brightness}
\label{radial}

We measured the surface brightness, $\Sigma(\theta)$, as a function of the angular radius, $\theta$, using a concentric set of  annular apertures, each 0.2\arcsec~wide and extending to 22\arcsec~radius.
For this purpose, we used the composite image on UT 2019 October 12 (6240 s duration from four HST orbits), giving the highest signal-to-noise ratios in the data.
Background signal was determined using  a concentric annulus extending from 22\arcsec~to 26\arcsec,  considering the limited WFC3 field of view with cut edges.
Although the faint edge of the tail extends $\sim$40\arcsec~from the nucleus,
for most azimuth angles, the optically dominant portion of the coma ($\lesssim$20\arcsec) is much smaller than 40\arcsec.
Thus, the tail was largely excluded by taking the median within the annulus.
The surface brightness profile on October 12 (Figure \ref{sbr}) shows significant curvature at $\sim$2\arcsec~from the nucleus and steepens at its outer edge.
In the central region, $\theta <$ 0.2\arcsec, the profile is affected by convolution with the point-spread function of HST, where the central region profile fitting is given in Jewitt et al.~(2020).
In the outer region ($\theta >$ 7\arcsec), uncertainties in the background subtraction become dominant.

The profile in the range 0.2\arcsec~$\le \theta \le$ 2\arcsec~range  is fitted by a power law, $\Sigma(\theta) \propto \theta^{m}$, with $m$ = -1.00$\pm$0.01.  This is very close to the $m = -1$ slope expected for the spherical, steady-state coma. 
In the 2\arcsec~$\le \theta \le$ 7\arcsec~range, we find a steeper $m$ = -1.50$\pm$0.02, consistent with the $m = -3/2$ slope caused by radiation pressure acceleration of the particles (Jewitt \& Meech~1987).
The location of the knee ($\sim$2\arcsec) was determined by optimizing the $m = -1$ and $m = -3/2$ fits.
The knee in the profile reflects the extent of the coma in the sunward direction, $X_R$, which is limited by the radiation pressure,

\begin{equation}
X_R = \frac{v^2}{2 \beta g_{\odot}}
\end{equation}

\noindent where $v$ is the speed of the ejected particles and $g_{\odot} \sin(\alpha) \approx g_{\odot}$ is the local solar gravitational acceleration.  Substituting $X_R = 4\times10^3$ km and $g_{\odot}$ = 10$^{-3}$ m s$^{-2}$, we find $v$ = 9 $(\beta/\beta_0)^{1/2}$ m s$^{-1}$ with $\beta_0$ = 0.01.

The  surface brightness gradient of the inner coma, $m$ = -1.00$\pm$0.01, suggests that the dust from 2I is released continuously and in steady-state (Jewitt \& Luu 2019).
The surface brightness of later dates was further measured, showing no significant change during the period from October 12 to January 03.
Although a slightly steeper profile was observed on January 29, this observation was made in a crowded star field and the background uncertainty increased significantly.  We consider the coma profiles from all epochs as being broadly consistent.

\smallskip
\subsection{Photometry}

We obtained photometry from each composite image (Figure \ref{images}) using a set of six circular apertures having fixed radii from 500 to 16,000~km, when projected to the distance of 2I.
The sky background was determined within a concentric annulus with inner and outer radii of 22\arcsec~and 26\arcsec, respectively (cf. Section \ref{radial}).
Flux calibration was performed using the online WFC3 Exposure Time calculations for a G2V source in the F350LP filter.
We converted the apparent magnitudes, $V$, to absolute magnitudes, $H$, using

\begin{equation}
H = V - 5\log_{10}(r_H \Delta) - f(\alpha)
\end{equation}

\noindent in which $r_H$ and $\Delta$ are the heliocentric and geocentric distances, respectively. $f(\alpha)$ is the phase function at solar phase angle $\alpha$.
In the absence of an empirical determination, we used $f(\alpha) = 0.04\alpha$, consistent with Jewitt et al.~(2020). More elaborate phase functions were considered in Hui et al.~(2020) and Ye et al.~(2020) but these are observationally unconstrained.

The resulting absolute magnitude is converted into the effective scattering cross-section, $C_e$ [km$^2$], by
 
\begin{equation}
C_e = \frac{1.5\times 10^6}{p_V} 10^{-0.4 H}
\end{equation}

\noindent where $p_V$ is the geometric albedo.  We assume $p_V$ = 0.1, consistent with the albedos of solar system cometary dust (Zubko et al.~2017).
For each date and aperture radius, $V$, $H$, and $C_e$ are summarized in Table \ref{phot}.

\begin{figure}
\epsscale{1.0}
\plotone{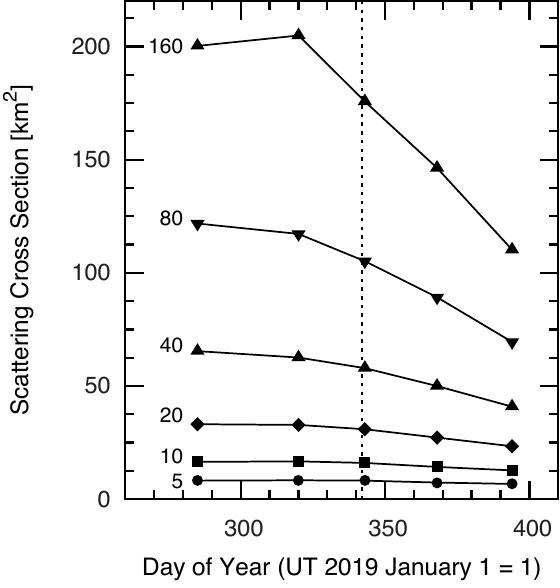}
\caption{Scattering cross-section as a function of time, expressed as Day of Year (DOY=1 on UT 2019 January 1).
The radii of the apertures (in units of 10$^2$ km) are marked.   The dotted line indicates the date of perihelion, UT 2019 December 08.\\ \label{ce}}
\end{figure}

In Figure \ref{ce}, the scattering cross-section within the 16,000~km radius aperture peaks in November (1 month before perihelion), then declines until the end of January by about a factor of two.
On the other hand, the scattering cross-section within the central aperture remains relatively constant.

\begin{deluxetable*}{lcccccc}
\tabletypesize{\scriptsize}
\tablecaption{Photometry with Fixed Linear Radius Apertures\tablenotemark{a}
\label{phot}}
\tablewidth{0pt}

\tablehead{ \colhead{UT Date} & 500 km & 1000 km  & 2000 km & 4000 km & 8000 km & 16000 km}
\startdata
2019 Oct 12 & 20.57/15.66/8.18 & 19.81/14.90/16.5 & 19.05/14.14/33.1 & 18.31/13.40/65.4 & 17.64/12.73/121.8 & 17.10/12.19/200.2\\
2019 Nov 16 & 19.99/15.63/8.36 & 19.24/14.88/16.7 & 18.50/14.15/32.8 & 17.80/13.45/62.6 & 17.12/12.77/117.2 & 16.52/12.16/204.8\\
2019 Dec 09 & 19.80/15.65/8.22 & 19.07/14.93/16.0 & 18.36/14.22/30.9 & 17.68/13.53/57.9 & 17.03/12.89/105.1 & 16.47/12.33/175.7\\
2020 Jan 03 & 19.93/15.77/7.36 & 19.20/15.05/14.3 & 18.51/14.35/27.2 & 17.85/13.69/50.0 & 17.22/13.06/89.2 & 16.68/12.53/146.3\\
2020 Jan 29 & 20.26/15.87/6.74 & 19.57/15.18/12.7 & 18.91/14.52/23.4 & 18.31/13.91/40.9 & 17.73/13.33/69.5 & 17.23/12.83/110.2

\enddata

\tablenotetext{a}{The Table lists the apparent magnitude, $V$, the absolute magnitude, $H$, and the scattering cross-section, $C_e$ [km$^2$], in the order $V/H/C_e$, for each of six photometry apertures.\\}

\end{deluxetable*}

\smallskip
\subsection{Perpendicular Profile}
\label{profile}

\begin{figure}
\epsscale{1.1}
\plotone{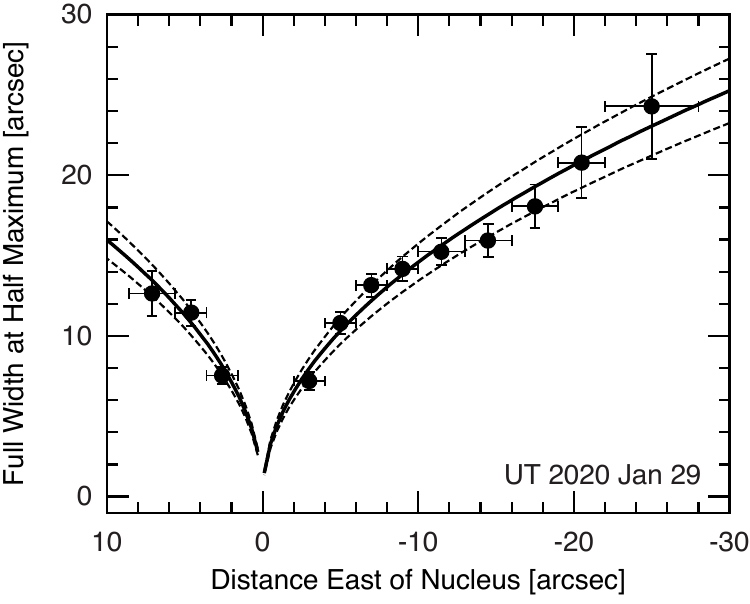}
\caption{FWHM of the dust tail as a function of the angular distance from the nucleus, observed at plane-crossing on UT 2020 January 29.
Horizontal bars indicate the width of the segment used to make the profiles, while vertical error bars denote uncertainties in the  FWHM measurement.
Best-fit lines (Equation \ref{width}) to the east of nucleus indicate ejection velocities 6.9$\pm$0.5 m s$^{-1}$ and to the west 6.3$\pm$0.5 m s$^{-1}$.\\ \label{fwhm}}
\end{figure}

Observations on UT 2020 January 29 were taken as the Earth passed through the projected orbit plane of 2I, and offer a particularly powerful constraint on the out-of-plane distribution of dust.  
Figure \ref{fwhm} shows the width of the tail, $w_T$, measured as a function of the projected angular distance from the nucleus, $\theta$. We measured the width from the FWHM of a series of surface  profiles cut perpendicular to the tail, in  2--6\arcsec~wide segments, where the tail gradually widened as the distance increased.

The width of the tail, $w_T$, is related to the distance from the nucleus, $\ell_T$, by

\begin{equation}
w_T = V_{\perp} \left(\frac{8 \ell_T}{g_{\odot}}\right)^{1/2}
\label{width}
\end{equation}

\noindent  where $V_{\perp}$ is the ejection velocity normal to the orbit plane and $g_{\odot}$ is the local solar gravitational acceleration (Jewitt et al.~2014).  
For simplicity, we assume $\ell_T \approx \theta$ and neglect projection effects.
We show Equation (\ref{width}) fitted to the data in Figure \ref{fwhm}, finding $V_{\perp}$ = 6.9$\pm$0.5 m s$^{-1}$ on the trail to the east of the nucleus, and $V_{\perp}$ = 6.3$\pm$0.5 m s$^{-1}$ to the west.  
Within the uncertainties, we take $V_{\perp} \sim$ 6.5 $(a/a_0)^{-1/2}$ m s$^{-1}$  as the dust ejection velocity,
where $a_0$ is the particle radius applicable to the measured trail width.  
We assume the optically dominant particle radius of $a_0=100~\mu$m (Jewitt \& Luu 2019).
These measurements are consistent with $V \sim$ 9 m s$^{-1}$ at $a \sim$ 100 $\mu$m inferred from the surface brightness.

\begin{figure}
\epsscale{0.96}
\plotone{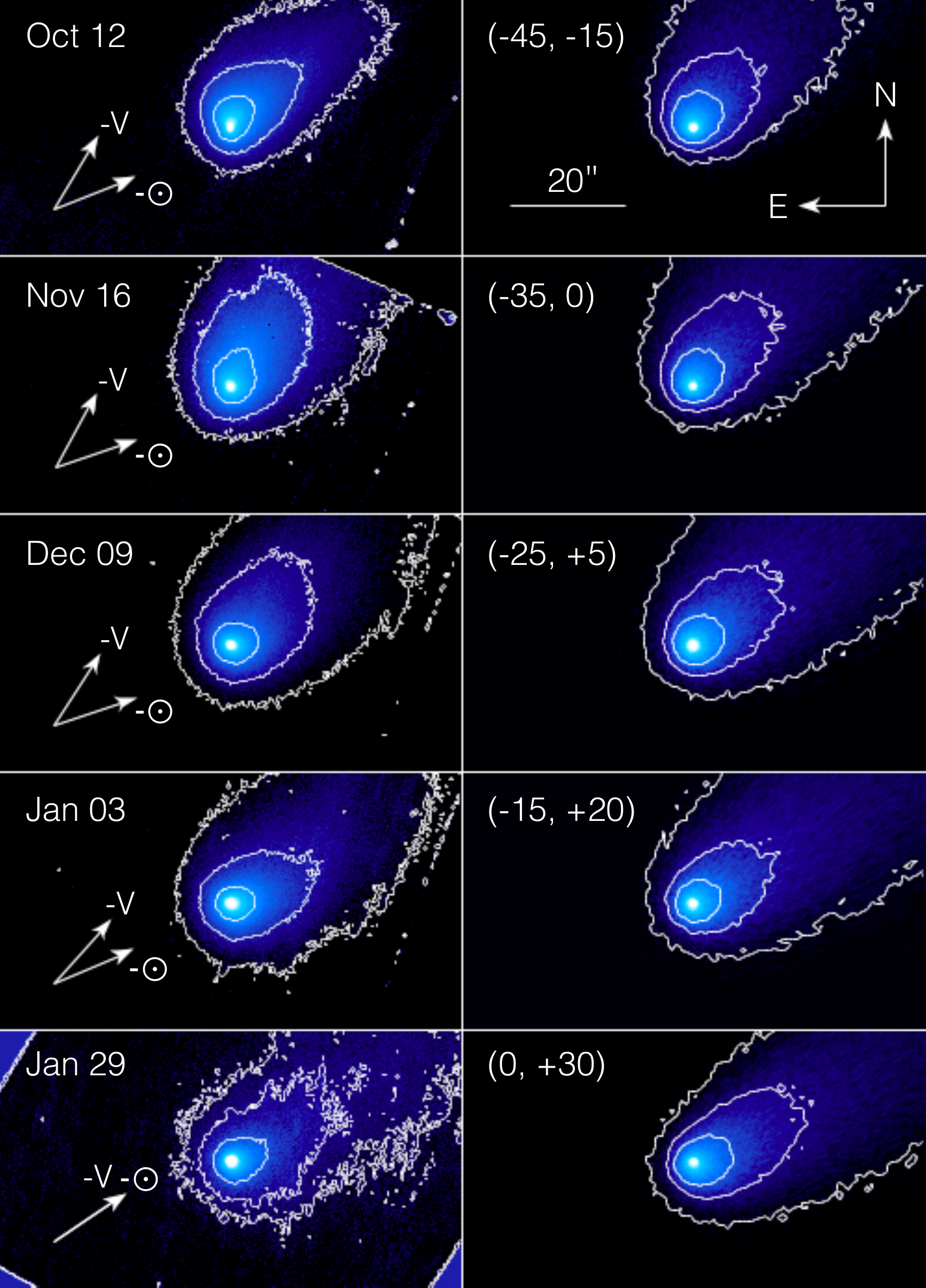}
\caption{Comparison between HST images (left) and Monte Carlo models (right) at five epochs of observation.
The best-fit direction of the dust jet axis for each date of observation, $(\alpha_{\rm jet}, \delta_{\rm jet})$, is varied, as marked.   A scale bar, the cardinal directions and the projected anti-solar direction ($-\odot$) and the negative heliocentric velocity vector ($-V$) are indicated.\\ \label{model}}
\end{figure}

\medskip
\section{DISCUSSION}

\subsection{Anisotropic Jet Model}
\label{aniso}

Inspection of the data  shows that the dust emission from 2I was anisotropic (Figure \ref{images}).
To explore this anisotropy, we conducted a series of  simulations of dust particle dynamics taking into account both solar gravity and radiation pressure. Our treatment of anisotropic dust emission is similar to that considered in Sekanina~(1987) but uses a Monte Carlo technique (developed by Ishiguro et al.~2007, Kim et al.~2017) in order to more easily explore the parameter space.

We consider  a dust jet  whose  axis is oriented perpendicular to the surface of a rotating, spherical nucleus.  Rotation of the nucleus sweeps the jet axis around a small circle, centered on the pole (whose celestial coordinates are ($\alpha_{\rm pol}$, $\delta_{\rm pol}$)) with a half-angle, $w$, equal to the colatitude of the source region.  The rotational  sweeping of the jet axis describes a conical shell of dust emission, whose axis lies parallel to the nucleus rotation axis (e.g.~Hsieh et al.~2011, see Figure~10).  We further assume that the jet is active only when its source region on the nucleus is illuminated by the Sun, so that the conical shell is only partially filled.
For the simulations discussed here, we assume $w$ = 45\degr. 
The jet contributes only a fraction of the total mass loss from 2I, with the bulk coming isotropically from the whole nucleus.  Indeed, measurements show that 2I has a large active fraction; with an upper limit to the nucleus radius $r_n \le$ 0.5 km and mass loss equivalent to  sublimation from a 0.4 km radius body (Jewitt et al.~2020, Hui et al.~2020), the active fraction must be $f_A \gtrsim (0.4/0.5)^2 \sim 0.6$.

The dust terminal ejection speed was assumed to be $V = V_0 (a/a_0)^{-1/2}$,
where $V_0$ is the ejection velocity of particles with $a_0=100~\mu$m.
In our model, the tail width is controlled by $V_{\perp} = V_0 \sin(w)$.
Substituting $V_{\perp}$ = 6.5 m~s$^{-1}$ (Section \ref{profile}) and $w$~=~45\degr, we adopt $V_0$ = 9.2 m~s$^{-1}$.
The choice of some of the parameters (jet colatitude $w$, power-law size index $q$, particle maximum size $a_1$) was guided by the prior application of this model to solar system comets (Ishiguro et al.~2007). We assumed that the ejected particles follow a differential power-law size distribution with index $q$ = -3.5, minimum particle radius $a_0=100~\mu$m (Jewitt \& Luu 2019), and maximum particle radius $a_1=1$~cm.
We assumed a dust production rate $\propto r_H^{-2}$, where $r_H$ is the  heliocentric distance.

Dust ejection is assumed to begin in 2019 June when $r_H \sim$ 4.5 AU (Jewitt \& Luu 2019). Ye et al.~(2020) report that 2I was weakly active at larger distances but we found, by trial and error,  that the results are not strongly dependent on the assumed starting date.  Specifically,  models with activity beginning in 2018 December (Ye et al.~2020)  generate  morphologies similar to those obtained with the later starting date.
Using the parameters as described above, we find plausible solutions for the projected direction of the dust jet axis, $(\alpha_{\rm jet}, \delta_{\rm jet})$, for each of the dates of observation. 
In this model, we assume that the effective time-averaged jet axis simply corresponds to the rotational pole ($\alpha_{\rm pol}$, $\delta_{\rm pol}$).
We created a number of model images using a wide range of jet directions having a 5\degr~interval.  The resulting model images were visually compared to the observations to find plausible solutions, and then we used least-squares fitting of tail isophotes  to find the best-fit solutions.
Figure \ref{model} compares the observations with the models on each date of observation.

After much experimentation, we could not find a unique set of fixed jet parameters to simultaneously match the morphology at all five  epochs of observation in Table \ref{geometry}.  Instead, the data are consistent with a time-dependent  jet  direction.
Several possibilities exist to account for a variable jet direction: 

\begin{enumerate}
\item The jet could be outgassing from an active area that is fixed on the surface of a nucleus whose rotation is itself precessing.  Precession, especially in small cometary nuclei like that of 2I, is an expected consequence of  outgassing torques caused by non-uniform sublimation.
\item  The nucleus rotation vector could be fixed, but the jet-producing active regions  could migrate over the surface with time.  This is natural on a nucleus having non-zero  obliquity, as previously unexposed latitudes are progressively brought into illumination by the Sun. 
\item The outgassing could be delayed into the local afternoon (``lagged'') because of the sluggish thermal response of the outgassing nucleus surface. Then, the projected direction of the jet would change depending on the viewing perspective.  This effect has long been known as a defining characteristic of comets (Whipple~1950).
\end{enumerate}

\begin{figure}
\epsscale{1.1}
\plotone{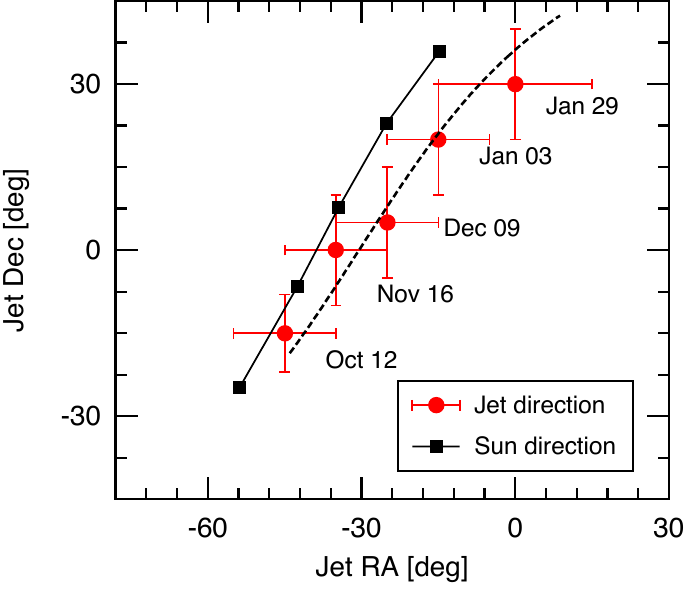}
\caption{The best-fit solutions for the projected direction of the dust jet axis, $(\alpha_{\rm jet}, \delta_{\rm jet})$, for each of the dates of observation (red circles).  Error bars indicate the range of plausible solutions.
The projected sunward directions for each date of observation (black squares), connected by the solid line, are plotted to guide the eye.  The dashed line marks the projected direction of a time-varying jet assuming thermal lag parameters $\theta_z$ = -15\degr~and $\phi_x$ = 10\degr.\\  \label{jet}}
\end{figure}

In Figure \ref{jet}, we plot the solutions for $(\alpha_{\rm jet}, \delta_{\rm jet})$ (red circles) for each date of observation. The figure shows a systematic drift in the model jet direction   with time from 2019 October to 2020 January.
Most interestingly, we find that the solutions follow the changing projected  direction of the Sun (black squares and solid line), but  are offset from it by $\Delta\theta\lesssim20\degr$.  As we argue below, the existence of this offset is most consistent with explanation (3), above, namely a thermal lag on the rotating nucleus, with peak mass loss occurring in the comet nucleus afternoon.

\bigskip
\subsection{Thermal Lag Model}
\label{lag}

If the outgassing rate is a measure of instantaneous local insolation, the sublimation rate should reach its maximum at meridian-crossing (local noon), and the projected axis of the coma should be centered on the Sun.
In  solar system comets, however, the activity does not generally occur at local noon due to a thermal lag on the rotating nucleus (Sekanina~1981).
Hence, we modified the anisotropic jet model to a ``thermal lag model,'' in which we assume that the direction of a time-varying jet $(x_{\rm jet}(t), y_{\rm jet}(t), z_{\rm jet}(t))$ can be obtained by rotating the direction to the Sun $(x_{\odot}(t), y_{\odot}(t), z_{\odot}(t))$ by $\theta_z$ around the $z$-axis and then by $\phi_x$ around the $x$-axis:

\begin{equation}
\begin{bmatrix}
        x_{\rm jet} \\
        y_{\rm jet} \\
        z_{\rm jet} \\
\end{bmatrix}
=
\begin{bmatrix}
        1 & 0 & 0\\
        0 & \text{cos($\phi_x$)} & \text{sin($\phi_x$)} \\
        0 & \text{-sin($\phi_x$)} & \text{cos($\phi_x$)} \\
    \end{bmatrix}
    \begin{bmatrix}
        \text{cos($\theta_z$)} & \text{sin($\theta_z$)} & 0\\
        \text{-sin($\theta_z$)} & \text{cos($\theta_z$)} & 0 \\
        0 & 0 & 1 \\
    \end{bmatrix}
    \begin{bmatrix}
        x_{\odot} \\
        y_{\odot} \\
        z_{\odot} \\
    \end{bmatrix}
    \label{rotation}
\end{equation}

\noindent where only two rotation angles are used to minimize the number of free parameters.
The three-dimensional vectors are expressed in heliocentric ecliptic coordinates.

We used the thermal lag model to find plausible rotation angles to match the observed data. 
Specifically, we simulated the sky-plane trajectories of the time-varying jet  for given ($\theta_z$, $\phi_x$)  from UT 2019 October 1 to 2020 February 1, and fitted them to the  measured  projected direction of the dust jet axis (Figure \ref{jet}).
We obtained plausible solutions for $-25\degr\lesssim\theta_z\lesssim-10\degr$ and $5\degr\lesssim\phi_x\lesssim15\degr$.
In Figure \ref{jet}, we show the best-fit simulated trajectories with $\theta_z=-15\degr$ and $\phi_x=10\degr$ (dashed line).  

The magnitude of the lag angle, $\Delta \theta = 20\degr$, is determined by the ratio of the thermal response time, $\tau$ (itself controlled by the thermal diffusivity of the nucleus and the depth  from which sublimated volatiles emanate) to the nucleus rotational period, $P$.  We write $\Delta\theta = 2\pi \tau/P$, with $\Delta \theta$ expressed in radians. The thermal response time for ice buried beneath a porous mantle of diffusivity $\kappa$ at depth $\ell$, is $\tau \sim \ell^2/\kappa$ so that

\begin{equation}
\ell = \left(\frac{\kappa \Delta\theta P}{2\pi}\right)^{1/2}
\label{depth}
\end{equation}

The median rotational period determined for solar system comets is $P$ = 6 hours (Kokotanekova et al.~2017).   The period is unlikely to be different from 6 hours by more than a factor of a few and therefore can have little effect on our estimate of $\ell$ through Equation (\ref{depth}).   Cometary materials have very low thermal diffusivities as a result of their porous structure.  As an example, we take $\kappa = 10^{-8}$ to $10^{-9}$ W m$^{-1}$ K$^{-1}$ to find from Equation (\ref{depth}) $\ell \sim$ 1 to 3 mm. Unless the diffusivity is orders of magnitude larger than we have assumed (which would imply a consolidated, dense structure unlike that of solar system comets) we must conclude that sublimation from 2I proceeds from ice either exposed at the surface or protected from it only by the thinnest of refractory mantles.

The thermal lag model is supported by the profiles of the CO rotational lines shown in ALMA spectra (Cordiner et al.~2020).  The lines show a slightly blueshifted component proving that CO is released from the source primarily on the illuminated side.
To fit the blueshifted profile, Cordiner et al.~(2020) assumed asymmetric outgassing, consistent with the thermal lag discussed here.

\smallskip
\subsection{Dust Production Rates}

Given the dust size distribution index $q$ = -3.5, the cross-section weighted mean particle radius contributing   most strongly  to the scattered light is given by $\overline{a} = (a_0 a_1)^{1/2}$, where $a_0$ and $a_1$ are minimum and maximum particle radii, respectively. 
To better estimate $a_0$ and $a_1$, we again used the thermal lag model (Section \ref{lag}) using a wide range of particle sizes.
We found that small particles ($a \lesssim 100~\mu$m) fail to produce  the observed  asymmetry in the coma, while the results are not strongly dependent on the particle maximum size.
The best-fit parameters from the model indicate that $a_0 \sim$ 100 $\mu$m and $a_1 \sim$ 0.1 to 10~cm.
We take $a_0$ = 100 $\mu$m and $a_1=1$~cm, yielding a surprisingly large $\overline{a}$ = 1 mm.
This value is 10 times larger than found by Jewitt \& Luu (2019),  based on order-of-magnitude considerations and more limited data and 10$^3$ times larger than the canonically assumed micron grain size in comets.
Millimeter-sized particles were independently identified in a model by Cremonese et al.~(2020).

We estimate an order of magnitude dust production rate using

\begin{equation}
\frac{dM}{dt} = \frac{4}{3}\frac{\rho \overline{a} C_e}{\tau_r},
\label{dmbdt}
\end{equation}

\noindent where $\rho$ = 500 kg m$^{-3}$ is the assumed particle density (Groussin et al. 2019),  $\overline{a}$ is the mean particle radius, $C_e$ is the scattering cross-section in a photometric aperture (Table 2), and $\tau_r$ is the residence time in the aperture.
The cross-section within the $\ell$ = 16,000~km radius aperture is $\sim$200 km$^2$ in 2019 October--November.
The residence time is given by $\tau_r = \ell/V$, where we take the ejection speed $V \sim 9$ m s$^{-1}$ (Section \ref{aniso}) to find $\tau_r \sim$ 2$\times 10^6$~s.
With $\overline{a}$ = 1 mm, Equation (\ref{dmbdt}) gives dust production rates $dM/dt =$  70 kg s$^{-1}$ in 2019 October--November declining to $\sim$35 kg s$^{-1}$ by 2020 January.
Given the many uncertainties (e.g.~in particle size (especially $a_1$), the power-law index ($q$)) the derived mass loss rate is probably not better than an order of magnitude estimate.
For comparison, the water production rates measured on UT 2019 October 11, based on spectroscopic detections of the [OI] 6300\AA~emission, were $dM/dt \sim$ 20 $\pm$ 5 kg s$^{-1}$ (McKay et al.~2020).  The  post-perihelion production rates of CO average $dM/dt \sim$ 20 to 40 kg s$^{-1}$ (Bodewits et al.~2020, Cordiner et al.~2020), indicating that 2I has a dust-to-gas ratio of order unity.

Assuming a  production rate of $dM_d/dt \sim$ 35 kg s$^{-1}$ sustained for 2$\times 10^7$ s (from the time of discovery to the latest observations),
we estimate the total mass loss from 2I of $\Delta M_d$ = 7$\times 10^8$ kg.  This corresponds to a shell of thickness $\Delta r_n = \Delta M_d /(4\pi r_n^2 \rho)$ on a spherical nucleus of radius $r_n$ and density $\rho$.  With $r_n$ = 500 m and $\rho$ = 500 kg m$^{-3}$, we compute $\Delta r_n \sim$ 0.4 m, about 100 times thicker than the  mantle thickness estimated from the phase lag.  We conclude that ices currently sublimating from 2I originate from beneath the original (pre-entry) surface of this object.   However, the layer thickness $\Delta r_n$ is comparable to, or  smaller than, the meter-thick layer likely to have been heavily processed by exposure to cosmic rays when in interstellar space (Cooper et al.~2003).  Thus, we conclude that the gases emanating from 2I are not necessarily pristine, but could have been chemically altered by prolonged exposure in the interstellar medium.

\smallskip
\subsection{Pole Orientation}

Cometary spin influences the outgassing rate through both diurnal and seasonal effects.
The success of our thermal lag model (Section \ref{lag}) suggests that the anisotropy of the mass loss is consistent with thermal lag on the rotating nucleus.
Based on this scenario, we estimate the rotation pole orientation of 2I as follows, in order to understand long-term variations in the coma brightness that may be influenced by seasonal effects.

Assuming that the nucleus spin is stable and peak mass loss occurs at subsolar latitude $\beta_{\odot}(t)$ and in the comet nucleus afternoon, we deduce the rotational pole ($\bf{r}_{\rm rot}$) that satisfies

\begin{equation}
\alpha_1(t) = \alpha_2(t) = 90\degr - \beta_{\odot}(t)
\label{pole}
\end{equation}

\noindent where $\alpha_1(t)$ is the angle between the pole ($\bf{r}_{\rm rot}$) and the comet-Sun vector (${\bf r}_{\odot}(t)$)
and $\alpha_2(t)$ is the angle between the pole ($\bf{r}_{\rm rot}$) and the direction of peak mass loss (${\bf r}_{\rm jet}(t)$).
Since ${\bf r}_{\rm jet}(t)$ and ${\bf r}_{\odot}(t)$ are related (Equation \ref{rotation}), ${\bf r}_{\rm rot}$ is uniquely determined once we fixed two Euler angles.
With $-25\degr\lesssim\theta_z\lesssim-10\degr$ and $5\degr\lesssim\phi_x\lesssim15\degr$ (Section \ref{lag}), we find plausible solutions for the pole orientation of $190\degr\lesssim\alpha_{\rm pol}\lesssim240\degr$ and $32\degr\lesssim\delta_{\rm pol}\lesssim66\degr$.
We take $\theta_z=-15\degr$ and $\phi_x=10\degr$ (Figure \ref{jet}) to find a best-fit pole orientation of $\alpha_{\rm pol}=$ 205$\degr$ and $\delta_{\rm pol}=$ 52$\degr$.  The implied nucleus obliquity is $\varepsilon$ = 30\degr.
Our best-fit pole is inconsistent with the pole solution found in Ye et al.~(2020), (i.e. either (340$\degr$, +30$\degr$) or (205$\degr$, -55$\degr$)), where they used a non-gravitational model fitted to the astrometric data to constrain the pole orientation. 
While our estimate of the rotational pole is most affected by the dust jet, the non-gravitational pole is largely controlled by the outgassing (e.g.~the gas jet).
We have no evidence regarding the correlation between dust jets and gas jets at 2I, which could cause differences in the estimated pole orientations.

\begin{figure}
\epsscale{1.15}
\plotone{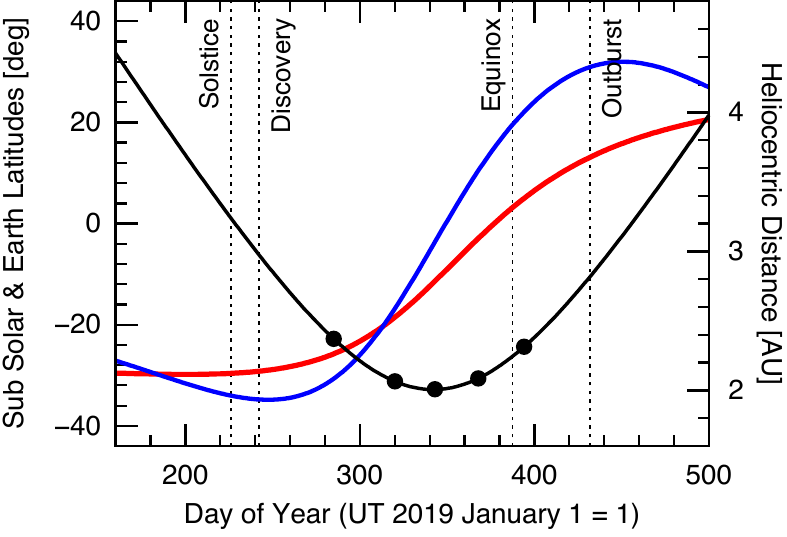}
\caption{Sub-solar (red) and Sub-Earth (blue) latitude of 2I as a function of time, together with the heliocentric distance (black solid line) on the right axis.  We assumed a rotation pole orientation of $\alpha_{\rm pol}=$ 205$\degr$ and $\delta_{\rm pol}=$ 52$\degr$.   Vertical lines indicate the dates of the discovery, southern solstice, equinox, and reported outburst.  Filled circles mark the HST visit dates.\\ \\ \label{season}}
\end{figure}

Figure \ref{season} shows the subsolar and sub-Earth latitudes of 2I as a function of time.
It is interesting to note that the southern solstice occurred in 2019 August, very close to the time of discovery. 
Between 2019 August and 2020 January, the subsolar latitude varied from $\beta_{\odot} \sim -35\degr$~to $\beta_{\odot} \sim 0\degr$, suggesting that long-term variations in the coma brightness and activity level may be influenced by seasonal effects.
For example, newly-reported photometric outbursts (Drahus et al.~2020) and the release of a fragment (Jewitt et al.~2020b) could  result from a seasonal effect, as  a previously unexposed and ice-rich region in the northern hemisphere turned into sunlight.  Our model shows that, as 2I receeds from the Sun, illumination of the previously unexposed hemisphere will grow.

\section{SUMMARY}

We studied the dust coma of interstellar comet 2I/Borisov on five occasions between UT 2019 October 12 (heliocentric distance $r_H$ = 2.370 AU, inbound) and 2020 January 29 ($r_H$ = 2.313 AU, outbound).  Our   Hubble Space Telescope data confirm and extend earlier reports that the dust from 2I/Borisov  is released continuously and in steady-state, as indicated by the logarithmic surface brightness gradient of the inner coma being close to $m = -1$.  The effective coma particle radius is a surprisingly large $\overline{a} \sim$ 1 mm.  We obtain the following new results;

\begin{enumerate}

\item 
Particle ejection velocities measured normal to the orbit plane are $V_{\perp} \sim$ 6.5 $(a/a_0)^{-1/2}$ m s$^{-1}$, where $a_0 = 100 \mu$m.    The $a^{-1/2}$ functional dependence is consistent with  particle ejection by  gas drag, as suggested independently by spectroscopic detections of gas. 

\item The total dust mass ejected between 2019 August and 2020 January corresponds to loss of a surface shell on the nucleus only $\sim$0.4 m thick.  This shell is susceptible to past chemical processing in the interstellar medium by cosmic rays, meaning that the ejected materials cannot necessarily be considered as pristine.

\item Persistent asymmetry in the coma suggests a thermal lag on the rotating nucleus, causing peak mass loss to occur in the comet nucleus afternoon.  The magnitude of the lag implies the existence of, at most, a millimeter-thick refractory mantle on the nucleus.

\item We deduce  a best-fit pole orientation of $\alpha_{\rm pol} =$ 205$\degr$ and $\delta_{\rm pol} = $ 52$\degr$.
The nucleus obliquity is $\varepsilon =$ 30\degr.  The subsolar latitude varied from $\beta_{\odot} \sim$ -35\degr~at the time of discovery to $\beta_{\odot} \sim$ 0\degr~in 2020 January, suggesting the importance of seasonal effects.   Subsequent activity likely results from new regions activated as the northern hemisphere is  illuminated for the first time.

\end{enumerate}

\acknowledgments

We thank the anonymous referee for helpful comments on this manuscript.
Based on observations made under GO 16009 and 16041 with the NASA/ESA Hubble Space Telescope obtained from the Space Telescope Science Institute,  operated by the Association of Universities for Research in Astronomy, Inc., under NASA contract NAS 5-26555.   Y.K. and J.A. were supported by the European Research Council Starting Grant 757390 ``CAstRA".

{\it Facility:}  \facility{HST (WFC3)}.

\end{document}